\documentclass[11pt]{article}

\usepackage{natbib}
\usepackage{graphicx}
\usepackage{amssymb}
\usepackage{amsmath}
\usepackage[nolists]{endfloat}

\newcommand{\comment}[1]{}

\newcommand{\bfc}{\hbox{\boldmath$c$}}
\newcommand{\bfn}{\hbox{\boldmath$n$}}
\newcommand{\bfp}{\hbox{\boldmath$p$}}
\newcommand{\bfv}{\hbox{\boldmath$v$}}
\newcommand{\bfx}{\hbox{\boldmath$x$}}
\newcommand{\bfz}{\hbox{\boldmath$z$}}

\newcommand{\bfalpha}{\hbox{\boldmath$\alpha$}}
\newcommand{\bfpsi}{\hbox{\boldmath$\psi$}}

\newcommand{\logit}{\hbox{logit}}

\begin{document}

\begin{center}
\Large{Estimation of population size when capture probability depends on individual states}

\vspace{5 mm}

\normalsize
Hannah Worthington$^1$, Rachel S. McCrea$^2$, Ruth King$^3$ and Richard A. Griffiths$^4$\\
$^1$School of Mathematics and Statistics and Centre for Research into Ecological and Environmental Modelling, University of St Andrews, St Andrews, Fife, KY16 9LZ, UK \\
$^2$School of Mathematics, Statistics and Actuarial Science, University of Kent, Canterbury, Kent, CT2 7NF, UK\\
$^3$School of Mathematics, University of Edinburgh, Edinburgh, EH9 3FD, UK\\
$^4$Durrell Institute of Conservation and Ecology and School of Anthropology and Conservation, University of Kent, Canterbury, Kent, CT2 7NR, UK\\
\emph{hw233@st-andrews.ac.uk}
\end{center}

\begin{abstract}
We develop a multi-state model to estimate the size of a closed population from ecological capture-recapture studies.  We consider the case where capture-recapture data are not of a simple binary form, but where the state of an individual is also recorded upon every capture as a discrete variable.  The proposed multi-state model can be regarded as a generalisation of the commonly applied set of closed population models to a multi-state form.  The model permits individuals to move between the different discrete states, whilst allowing heterogeneity within the capture probabilities.  A closed-form expression for the likelihood is presented in terms of a set of sufficient statistics.  The link between existing models for capture heterogeneity are established, and simulation is used to show that the estimate of population size can be biased when movement between states is not accounted for.  The proposed unconditional approach is also compared to a conditional approach to assess estimation bias.  The model derived in this paper is motivated by a real ecological data set on great crested newts, \textit{Triturus cristatus}.
\end{abstract}


\section{Introduction}
The models presented within this paper focus on the estimation of the size of a closed population using real ecological capture-recapture data on a population of great crested newts \textit{Triturus cristatus}.  An assumption often made in the modelling of capture-recapture data is that of homogeneity in the probability of capture.  When estimating the size of a closed population, one in which the population being sampled remains constant across all capture occasions, ignoring this assumption can lead to biased estimates of abundance \citep{seber82, hwang05}.  A number of models have been proposed that relax this homogeneity assumption.  In particular, \citet{pollock74} and \citet{otis78} proposed a set of eight closed population capture-recapture models.  These models allow the probability of capture to be affected by three factors: time (capture probabilities vary by occasion); behaviour (probability of initial capture is different to all subsequent recaptures) and heterogeneity (each individual has a different capture probability).  These models have been fitted using a variety of methods including maximum likelihood \citep{otis78, agresti94, norris96, coull99, pledger00}, the jackknife \citep{burnham78, burnham79, pollock83}, moment methods based on sample coverage \citep{chao92} and Bayesian methods \citep{casteldine81, gazey86, smith88, smith91, george92, diebolt93, ghosh05, king08}.  To specifically address the problem of heterogeneity in the capture probabilities a variety of models have been proposed, including finite mixtures \citep{diebolt93, agresti94, norris96, pledger00} and infinite mixtures \citep{coull99, dorazio03}.  A comparison of examples of the two types of mixture through simulation are presented in \citet{pledger05} and \citet{dorazio05}.  An issue that commonly arises when estimating the size of closed populations is that different individual heterogeneity models which may be deemed to fit the data equally well can give rise to very different estimates of the abundance \citep{link03, link06}.  An extended mixture model which provides a convenient framework for model selection is presented in \citet{morgan08}; see also \citet{holzmann06}.

We extend the previous models for closed capture-recapture data to account for individual heterogeneity when the ``state'' of an individual is also recorded as a discrete variable.  In standard closed capture-recapture studies, for each capture occasion, individuals from a closed population are sampled, returned to the population, and on subsequent occasions attempts are made to recapture them.  Individuals within the population are marked when initially captured.  If the marking method used assigns a unique mark to each captured individual (e.g. individual ID tags, natural physical markings) then an individual encounter history can be constructed for each individual observed within the study.  This history typically takes the form of a vector of 0s and 1s: with a 0 denoting an individual was not encountered and a 1 denoting an individual was captured.  We consider the case where individuals may be observed in different discrete states. For example, states may correspond to ``resting'' and ``foraging'' or ``breeding'' and ``not breeding''. Observed histories then correspond to whether an individual is observed or not, and, given that an individual is observed, its corresponding state. For example, suppose that there are three states labelled 1, 2, and 3. The encounter history
\[
1 \: 0 \: 0 \: 3 \: 2 \: 0
\]
represents that the given individual was observed (and marked) on occasion 1 in state 1, observed on occasion 4 in state 3 and occasion 5 in state 2 and unobserved on occasions 2, 3 and 6. In general, individuals are able to move between the states during the study period and the capture probabilities may be dependent on the state of an individual. For example, if the states correspond to ``resting'' and ``foraging'', the capture probabilities may be very different with a significantly higher capture probability for individuals in the ``foraging'' state compared to ``resting''. Failure to account for the state dependence may result in biased population estimates.  To estimate the number of unobserved individuals (and hence the total population size) a model is fitted to the observed data, permitting the estimation of the associated model parameters and the number of unobserved individuals. For such models, a number of standard assumptions are made. These include the population as a whole is closed; marks can not be lost and individuals are identified without error (see \cite{mccrea14}, chapter 3, for further discussion and references therein).

The model we present can be considered a generalisation of the time-dependent multi-state closed population model of \citet{schwarz95} to a form that additionally includes trap-dependence and heterogeneity in the capture probabilities.  It may also be seen as a closed-form capture-recapture equivalent to the Arnason-Schwarz (AS) model \citep{arnason72, arnason73, brownie93, schwarz93, king03, lebreton09}, for open capture-recapture data.  Initially developed for multi-site capture-recapture data, but more generally applicable to individuals captured in discrete states, the AS model is a multi-state generalisation of the Cormack-Jolly-Seber (CJS) model \citep{cormack64, jolly65, seber65}.  The CJS and AS models allow for a time-dependence in the capture probabilities, with the AS model additionally able to allow capture probabilities to be state-dependent. However, these models condition on the first capture of an individual and so are unable to estimate the total population size directly.  \cite{dupuis07} consider a similar multi-state extension for the Jolly-Seber model for estimating abundance in open populations, fitted within a Bayesian (data augmentation) framework.

We consider a similar AS-type state-dependence in the closed population scenario, within an explicit closed-form likelihood framework where population size is estimated directly through the likelihood. In particular, we compare the performance of the new unconditional multi-state model to the existing single-state models and a conditional approach where the population size is not estimated directly through the likelihood.  We present the likelihood in terms of a set of minimal sufficient statistics which permits the fit of the model to be assessed using a Pearson chi-squared test.  

The motivation for developing this methodology relates to a study on great crested newts.  A species with protected status in Europe, individuals within the study population are captured weekly throughout the breeding season, with the additional state information referring to the pond in which the individuals are captured.  Originally consisting of four ponds, the study site was extended to a total of eight ponds in 2009 with the new ponds being first colonised in the 2010 breeding season.  How these new ponds have been colonised, the effectiveness of the traps to capture individuals and whether there are differences in the capture probabilities between the old and new ponds are of particular interest.  Ignoring any differences between the old and new ponds, for instance differing amounts of vegetation which may be affecting the probability of capture, may lead to poorer estimates of the total population size, and for this study the states themselves are of interest.

In Section~\ref{sec:sscmodels} we review the construction of existing single-state closed population models in terms of sets of sufficient statistics, before introducing the likelihood function for the multi-state model in Section~\ref{sec:msmodellik}.  The performance of the multi-state model is compared to a conditional approach and the existing heterogeneity models using simulation in Section~\ref{sec:simstudy} with a particular focus on the bias and precision of the population size estimates and the ability of the new model to estimate state specific parameters. The new model is applied to the data set of great crested newts in Section~\ref{sec:newts}.  The paper concludes with a general discussion in Section~\ref{sec:diss}.

\section{Single-state closed population models}\label{sec:sscmodels}

Consider a study with $T$ capture occasions labelled $t=1,\ldots,T$ and let $N$ denote the total population size, which is to be estimated.  Let the set of encounter histories be given by $\bfx = \{x_{it}: i=1,\ldots,N, \ t=1,\ldots,T\}$, where $x_{it} = 1$ indicates individual $i$ was captured on occasion $t$, and $x_{it} = 0$ indicates individual $i$ was not captured on occasion $t$.  We let $n$ denote the number of observed individuals within the study. Further, we define the set of capture probabilities $\bfp = \{p_{it}: i=1,\ldots,N, \ t=1,\ldots,T\}$ where $p_{it}$ denotes the probability individual $i$ is captured on occasion $t$ (we note that for generality the initial capture probability and recapture probabilities may be different).  The overall likelihood expression for a closed population model can be expressed in the form,
\begin{equation}\label{eq1}
L(\bfx ; N, \bfp) \propto \frac{N!}{(N-n)!} \prod_{i=1}^{N}{\hbox{Pr(encounter history for individual $i$)}}.
\end{equation}

The existing modelling approaches for data of this type differ by the capture probability parameter dependence.  Using the notation of \citet{otis78}, we denote models by $M_\gamma$ where $\gamma$ describes the dependence structure placed on the capture probabilities. In general, $\gamma \subseteq \{t, b, h\}$, where $t$ denotes temporal dependence; $b$ denotes behavioural dependence (or trap response); and $h$ denotes individual heterogeneity. This leads to a total of eight different model dependencies, corresponding to the inclusion/exclusion of each of the different types of dependence, with $M_0$ denoting the model with a constant capture probability which ignores all three dependencies. We note that given a particular form of heterogeneity, multiple models may be defined in terms of the specific dependence. For example, and of particular interest in this paper, a number of different models have been proposed to include individual heterogeneity. In particular, in the absence of additional individual covariate information, the capture probabilities have been specified as finite or infinite mixtures models. These include (with associated notation):
\begin{description}
	\item $M_{h(k)}$: individual capture probabilities come from a mixture model with $k$ components \citep{pledger00};
	\item $M_{h(be)}$: individual capture probabilities specified to be from an underlying beta distribution \citep{burnham72,dorazio03};
	\item $M_{h(b-be)}$: individual capture probabilities come from a mixture model with two components: one component simply has a fixed capture probability while the other component is specified to be from some underlying beta distribution \citep{morgan08}.
\end{description}

For a given model, the corresponding likelihood function can be specified and maximised to obtain the MLEs of the model parameters (including the beta distribution parameters for models $M_{h(be)}$ and $M_{h(b-be)}$). The likelihood function given above is a function of the observed encounter histories $\bfx$. However, this likelihood can be expressed more efficiently via the use of sufficient statistics for some of the models. In particular, for models $M_0$ and $M_h$ dependencies, a set of sufficient statistics is the Schnabel census \citep{schnabel38}, defined to be $\{f_1,f_2,\ldots,f_T\}$, where $f_j$ denotes the number of individuals captured on a total of $j$ occasions. The Schnabel census denotes the set of minimal sufficient statistics for the heterogeneity models $M_h$; for model $M_0$ the minimal sufficient statistics reduce to $f=\sum_{j=1}^{T}{j f_j}$, corresponding to simply the total number of encounters over the study. For model $M_t$ the minimal sufficient statistics are $\{n_1,\ldots,n_T\}$, where $n_t$ denotes the number of individuals captured on occasion $t=1,\dots,T$. For model $M_{tb}$ minimal sufficient statistics are given by $\{z_1,\dots,z_T,n_2,\dots,n_T\}$ where $z_t$ denotes the number of individuals captured for the first time on occasion $t$. Finally, for model $M_b$, the sufficient statistics can be reduced to $\{n, y, f\}$ where $n = \sum_{t=1}^T z_t$, corresponding to the number of observed individuals within the study; $y = \sum_{t=1}^T(t-1) z_t$, corresponding to total number of capture events before initial observation summed over all captured individuals; and $f = \sum_{t=1}^T n_t$, the total number of captures (equivalent to the equation for $f$ given above).

The use of sufficient statistics allows for an efficient evaluation of the likelihood. In addition, they have the advantage of being able to be used to assess the performance of each of these models through the calculation of the Pearson chi-squared statistic, since the likelihood of the data is of multinomial form.  The advantage of using sufficient statistics, compared to a bootstrap method, is that the likelihood and goodness-of-fit test require only one evaluation of the likelihood which may be advantageous for large populations.

\section{Multi-state closed population model}\label{sec:msmodellik}

We now extend the previous closed population models for standard encounter histories to those with individual time-varying discrete state information. In particular, we let ${\cal R}$ denote the set of possible discrete states, which for convenience we label as $r=1,\dots,R$. Following the AS analogy to the CJS model, we assume that movement between these states is modelled as a first-order Markov process. We then define the following model parameters:

\begin{description}
\item $p_t(r)$: the probability an individual is initially captured at time $t=1,\dots,T$ given that the individual is in state $r \in {\cal R}$ at this time;
\item $c_t(r)$: the probability an individual is recaptured at time $t=2,\dots,T$ given that the individual is in state $r \in {\cal R}$ at this time;
\item $\psi_t(r,s)$: probability an individual is in state $s \in {\cal R}$ at time $t+1$, given that an individual is in state $r \in {\cal R}$ at time $t=1,\dots,T-1$;
\item $\alpha(r)$: probability an individual is in state $r \in {\cal R}$ at time 1.
\end{description}

For notational convenience we let $\bfp = \{p_t(r): t=1,\dots,T, \ r \in {\cal R}\}$, $\bfc = \{c_t(r): t=2,\dots,T, \ r \in {\cal R}\}$, $\bfpsi = \{\psi_t(r,s): t=1,\dots,T-1, \ r,s \in {\cal R}\}$ and $\bfalpha = \{\alpha(r): r \in {\cal R}\}$. We note that by definition, $\sum_{r \in {\cal R}}{\alpha(r)} = 1$. To retain model identifiability the recapture probabilities are specified as a function of the initial capture probabilities, such that $\logit c_t(r) =  \logit p_t(r) + \beta$, where $\beta$ denotes the trap dependence; $\beta < 0$ corresponds to a trap shy response; and $\beta > 0$ a trap happy response (see for example \cite{chao01, king08} for the analogous single-state case).

\subsection{Likelihood formulation}

The likelihood function is again of the same form as given in equation (\ref{eq1}), where now the probability of the encounter history includes not only detection/non-detection at each time but also the associated discrete state. In order to evaluate the likelihood efficiently, we follow \citet{king14} and consider all possible partial encounter histories that could be observed corresponding to (i) the beginning of the study to initial capture; (ii) successive captures; and (iii) final capture to the end of the study. This leads to the following sufficient statistics:

\begin{description}
\item (i) $z_t(r)$: the number of individuals that are observed for the first time at time $t=1,\dots,T$ in state $r \in {\cal R}$;
\item (ii) $n_{t_1,t_2}(r,s)$: the number of individuals that are observed at time $t_1 =1,\dots,T-1$ in state $r \in {\cal R}$ and next observed at time $t_2 = t_1+1,\dots,T$ in state $s \in {\cal R}$;
\item (iii) $v_t(r)$: the number of individuals that are observed for the last time at time $t=1,\dots,T-1$ in state $r \in {\cal R}$.
\end{description}

\noindent For notational convenience we set $\bfz = \{z_t(r): t=1,\dots,T, \ r \in {\cal R}\}$, $\bfn = \{n_{t_1,t_2}(r,s): t_1 = 1,\dots,T-1, \ t_2 = t_1+1,\dots,T, \ r,s \in {\cal R}\}$ and $\bfv = \{v_t(r): t = 1,\dots,T-1, \ r \in {\cal R}\}$.

In order to express the likelihood as a function of the above sufficient statistics, we need to calculate the probabilities for each of the associated partial encounter histories. In deriving these probabilities, we consider similar notation to \citet{king14}. We let $Q_{t_1,t_2}(r,s)$ denote the probability that an individual in state $r \in {\cal R}$ at time $t_1=1,\dots,T-1$ is in state $s \in {\cal R}$ at time $t_2=t_1+1,\dots,T$ and not observed at any times between $t_1$ and $t_2$. The form of this probability is dependent on whether an individual has yet to be captured for the first time or has been previously captured on at least one occasion.  We let $Q_{t_1,t_2}^P(r,s)$ denote the former and $Q_{t_1,t_2}^C(r,s)$ the latter scenario.  (If the capture probabilities are not behaviour dependent this distinction is not required.)  Then it follows immediately that
\[
Q_{t_1,t_2}^P(r,s) = \left\{ \begin{array}{ll} \psi_{t_1}(r,s) & t_2=t_1+1 \\
\sum_{u \in \cal R}{\psi_{t_1}(r,u) (1-p_{t_1+1}(u)) Q_{t_1+1,t_2}^P(u,s)} & t_2 = t_1+2,\dots,T. \\
\end{array} \right.
\]

$Q_{t_1,t_2}^C(r,s)$ follows analogously using the appropriate recapture probabilities.  We now consider the probabilities associated with each of the above sufficient statistics. We begin by considering the probabilities associated with an individual being observed for the first time (i.e.~case (i) and statistic $\bfz$). Let $\zeta_t(r)$ denote the probability an individual is initially captured at time $t=1,\dots,T$ in state $r \in {\cal R}$. Then using a probabilistic argument we have,
\[
\zeta_t(r) = \left\{ \begin{array}{ll} p_1(r) \alpha(r) & t = 1 \\
p_t(r) \sum_{u \in \cal R}{\alpha(u) (1-p_1(u)) Q_{1,t}^P(u,r)} & t =2,\dots,T. \\
\end{array} \right.
\]

We now consider case (ii) (associated with statistic $\bfn$) and the probability of being recaptured, conditional on the previous capture time. Let $O_{t_1,t_2}(r,s)$ denote the probability an individual in state $r \in {\cal R}$ at time $t_1=1,\dots,T-1$ is next recaptured in state $s \in {\cal R}$ at time $t_2 = t_1+1,\dots,T$. Then, by definition,
\[
O_{t_1,t_2}(r,s) = Q_{t_1,t_2}^C(r,s) c_{t_2}(s).
\]

The final case (iii) (associated with statistic $\bfv$) considers the probability an individual is not observed again within the study, following their final capture. Let $\chi_t(r)$ denote the probability an individual in state $r \in {\cal R}$ at time $t=1,\dots,T-1$ is not observed again during the study.  Then, for all $r \in {\cal R}$ it follows that,
\[
\chi_t(r) = \left\{\begin{array}{ll}1 & t = T \\
\sum_{u \in \cal R}{Q_{t,T}^C(r,u) (1-c_T(u))} & t = 1,\dots,T-1. \end{array}\right.
\]
By definition an individual observed at the last capture time is clearly not able to be observed again within the study, i.e.~the probability it is not seen again is 1 (this means that we do not need to consider this term within the likelihood expression).

Finally, in order to permit estimation of the total population size, we let $\rho$ denote the probability an individual is not observed within the study. From the law of total probability (considering all possible states an individual may be in at the first and last capture time) it follows that
\begin{eqnarray}\label{rho}
\rho = \sum_{r,s \in \cal R}{\alpha(r) (1-p_1(r)) Q_{1,T}^P(r,s) (1-p_T(s))}.
\end{eqnarray}

The corresponding likelihood function, specified as a function of the sufficient statistics is of the form,
\begin{eqnarray*}
L(\bfn, \bfv, \bfz; N, \bfp, \bfc, \bfpsi, \bfalpha) & \propto & \frac{N!}{(N-n)!} \rho^{N-n} \prod_{t=1}^{T} \prod_{r \in \cal R} \left[ \zeta_t(r)^{z_t(r)} \right] \\
 & \times & \prod_{t=1}^{T-1} \prod_{r \in \cal R} {\left[ \chi_t(r)^{v_t(r)} \prod_{\tau=t+1}^{T} \prod_{s \in \cal R} O_{t,\tau}(r,s)^{n_{t,\tau}(r,s)} \right]}.
\end{eqnarray*}

The above likelihood allows for temporal effects, behavioural effects and individual heterogeneity (in the form of discrete covariates), which we represent notationally as $M_{tbh}^{R}$, where the superscript, $R$, denotes the number of discrete states. Sub-models can be derived by specifying restrictions on the model parameters. In particular the basic dependence structures can be described with
\begin{description}
\item $M_0^R$: $p_t(r) = c_t(r) = p$, for all $r \in {\cal R}$ and $t=1,\dots,T$;
\item $M_t^R$: $p_t(r) = c_t(r) = p_t$, for all $r \in {\cal R}$ and $t=1,\dots,T$;
\item $M_b^R$: $p_t(r) = p$ and $c_t(r) = c$, for all $r \in {\cal R}$ and $t=1,\dots,T$;
\item $M_h^R$: $p_t(r) = c_t(r) = p(r)$, for all $r \in {\cal R}$ and $t=1,\dots,T$;
\end{description}
with associated restrictions for models with multiple dependencies. We note that the case of heterogeneous capture probabilities is fully determined by the additional discrete state information where a capture probability is estimated for each discrete state and remains constant for each state across all capture occasions.

Evaluating the likelihood through the sufficient statistics uses recursions similar to those in hidden Markov models (HMMs) but in more efficient forms.  In the HMM framework the likelihood considers each individual encounter history in turn.  By using more efficient sufficient statistics we are able to reduce the number of operations required to calculate the likelihood.  This is achieved by using the probabilities associated to each of the sufficient statistics for multiple partial histories.

\subsection{Conditional and Unconditional approaches}

\citet{bishop75} classified population size modelling into both conditional and unconditional approaches.  The unconditional approach involves maximising the full likelihood, written as a function of the observed capture histories (or associated sufficient statistics) and the number of unobserved individuals to obtain an MLE of the total population size $N$. The conditional approach \citep{sanathanan72} involves maximising the conditional likelihood (conditional on the number of observed individuals) to obtain estimates of the capture probabilities. The population size is then estimated using a Horvitz-Thompson-like estimator; see \citet[pp.33--35]{mccrea14} for the single-state case and \citet{schwarz95} for the multi-state case (though here estimation of the total population was not of interest for the given study).  Specifically:
\begin{eqnarray}\label{condN}
\widehat{N} & = & \frac{n}{1-\widehat{\rho}},
\end{eqnarray}
where $\hat{\rho}$ is calculated using Equation~\ref{rho} but using the MLEs obtained using the conditional likelihood.  \citet{fewster09} demonstrated that, for a wide class of models, the difference between the population size estimates obtained from the conditional and unconditional approaches is of order 1. However the differences were large when capture probabilities included both a behavioural and heterogeneity effect, and in this case advocated the use of unconditional approaches.  Here we develop the multi-state unconditional approach.

\subsection{State uncertainty}

Within the model derivation we have assumed states are known with certainty. In practice this may not be the case and the likelihood can be extended to incorporate state-uncertainty akin to the approach adopted by \citet{king14}.  This involves introducing further parameters corresponding to state assignment probability terms. In the case where no states are known with certainty, the model reduces to a multi-event-type model \citep{pradel05} corresponding to a finite mixture model which allows for transitions between states.  Conditional on the observed number of individuals, these multi-event models can be fitted within E-SURGE \citep{choquet09} to estimate the model parameters (though this package does not have the associated Horvitz-Thompson-like estimator incorporated into it). We further note that the limiting case where no states are observed upon recapture and there are no transitions between states, the model reduces to the mixture models proposed by \citet{pledger00}, where $\bfalpha$ represents the associated mixture probabilities.

\section{Simulation study}\label{sec:simstudy}

We conduct a simulation study of the proposed $M_h^{R}$ model (i.e.~no temporal or behavioural effects but a constant capture probability for each discrete state). We compare the performance of fitting this true covariate model with the corresponding conditional approach and a range of alternative individual heterogeneity models which ignore the state covariate (models $M_{h(2)}$, $M_{h(3)}$, $M_{h(be)}$ and $M_{h(b-be)}$), and models that ignore the individual heterogeneity all together (models $M_0$, $M_t$ and $M_b$). Of particular interest is the bias and precision of the population estimates for the different models, especially those that do not account for state-dependence in the capture probabilities. For each simulation we assume that there are six encounter occasions ($T=6$) and a total population size of 100 ($N=100$).  We consider four different sets of parameter values for the simulation study, corresponding to different numbers of states ($R=2$ and $R=3$) and different sets of transition matrices.  We evaluate two scenarios corresponding to ``low'' mobility and ``high'' mobility between states. In particular, for $R=2$ low mobility corresponds to $\psi(1,2)=0.3$ and $\psi(2,1)=0.2$; high mobility to $\psi(1,2)=0.9$ and $\psi(2,1)=0.6$. The equilibrium distribution for the low and high mobility cases are the same, and we set the initial state distribution to be equal to this equilibrium distribution, such that $\alpha(1) = 0.4$ (and $\alpha(2) = 0.6$). Finally, for each of these cases, the capture probabilities for the different states are set to $p(1)=0.15$ and $p(2)=0.4$. For $R=3$, the transition matrices are specified to be:
\begin{eqnarray*}
\left( \begin{array}{ccc} 0.76 & 0.12 & 0.12 \\
0.1 & 0.8 & 0.1 \\
0.15 & 0.15 & 0.7 \\
\end{array} \right)
 &  &
\left( \begin{array}{ccc} 0.28 & 0.36 & 0.36 \\
0.3 & 0.4 & 0.3 \\
0.45 & 0.45 & 0.1 \\
\end{array} \right)
\end{eqnarray*}
for low and high mobility respectively. Again, these transition matrices are specified such that they have the same equilibrium distribution, and we set the initial state distribution to be equal to this equilibrium distribution, such that $\alpha(1) = 0.33$, $\alpha(2) = 0.4$ (and $\alpha(3) = 0.27$). The corresponding state-dependent capture probabilities are defined to be $p(1)=0.15$, $p(2) = 0.25$ and $p(3)=0.4$. For each set of parameter values we simulate a total of 100 datasets and fit the following models to the data: $M_0$, $M_t$, $M_b$, $M_{h(2)}$, $M_{h(3)}$, $M_{h(be)}$, $M_{h(b-be)}$, the true model $M_h^{R}$ and the conditional model $M_h^{R(c)}$.  In the conditional model $N$ is not estimated directly through the likelihood but is calculated using the Horvitz-Thompson-like estimator in Equation~\ref{condN}.

\begin{figure}[!htp]
\begin{center}
\includegraphics[width=0.35\textwidth]{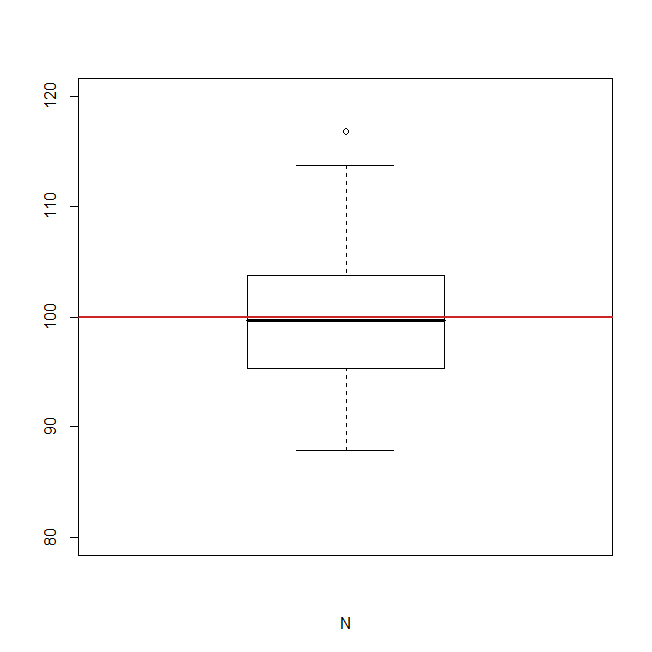}
\includegraphics[width=0.35\textwidth]{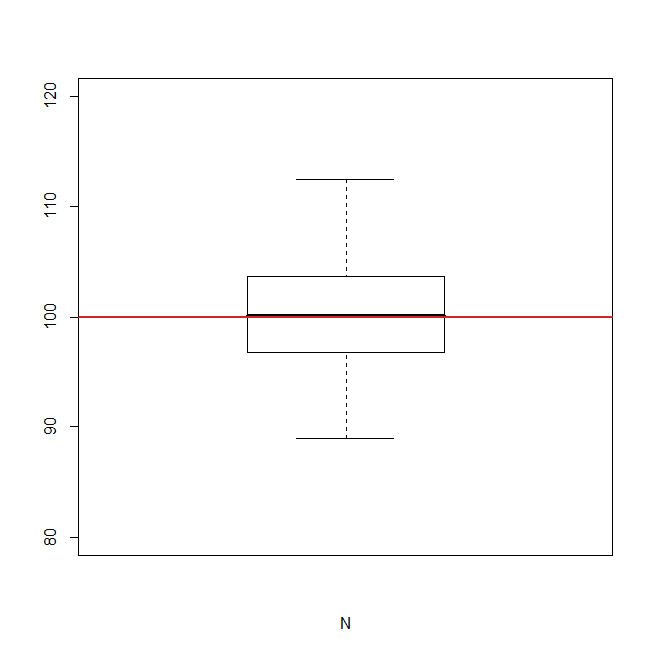}
\includegraphics[width=0.35\textwidth]{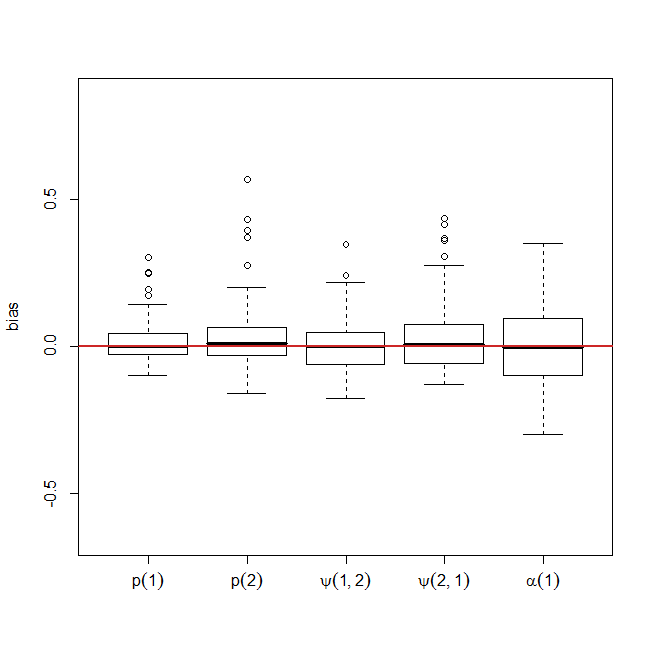}
\includegraphics[width=0.35\textwidth]{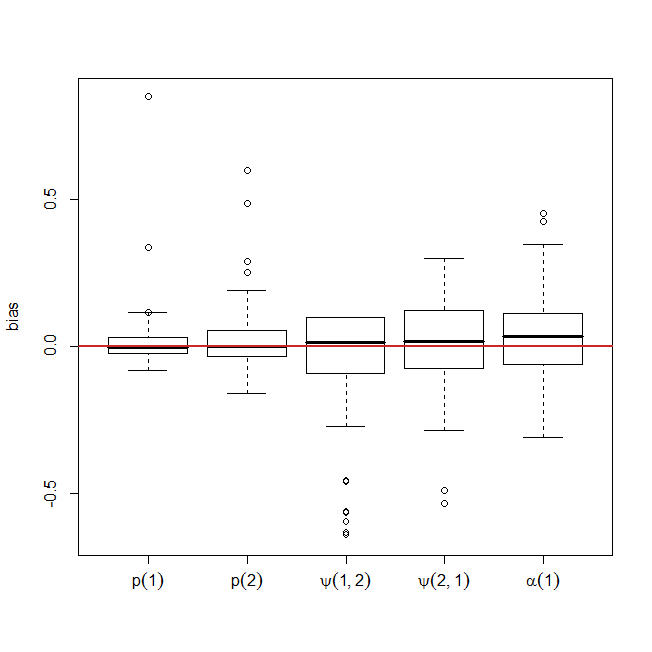}
\caption{Estimates of population size and bias of remaining model parameters for model $M_h^2$ for cases of low mobility (left) and high mobility (right). Parameter values used are given in the text.}
\label{fig:twostates}
\end{center}
\end{figure}

\begin{figure}[!htp]
\begin{center}
\includegraphics[width=0.35\textwidth]{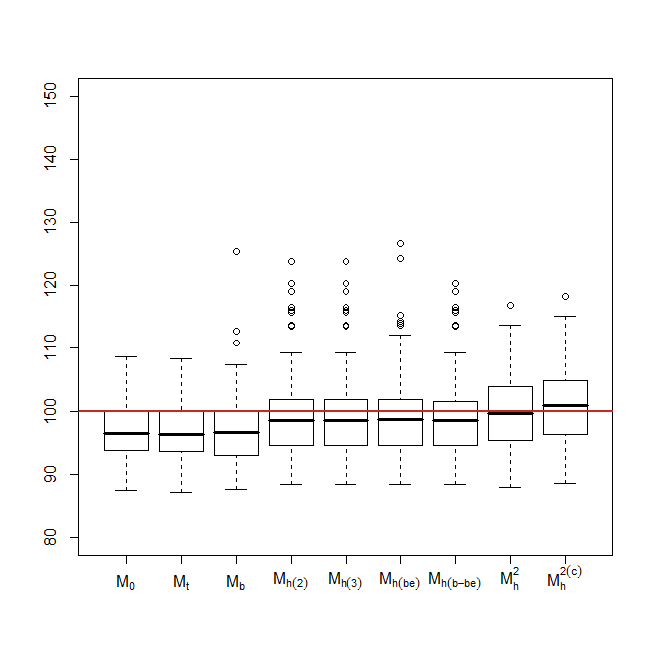}
\includegraphics[width=0.35\textwidth]{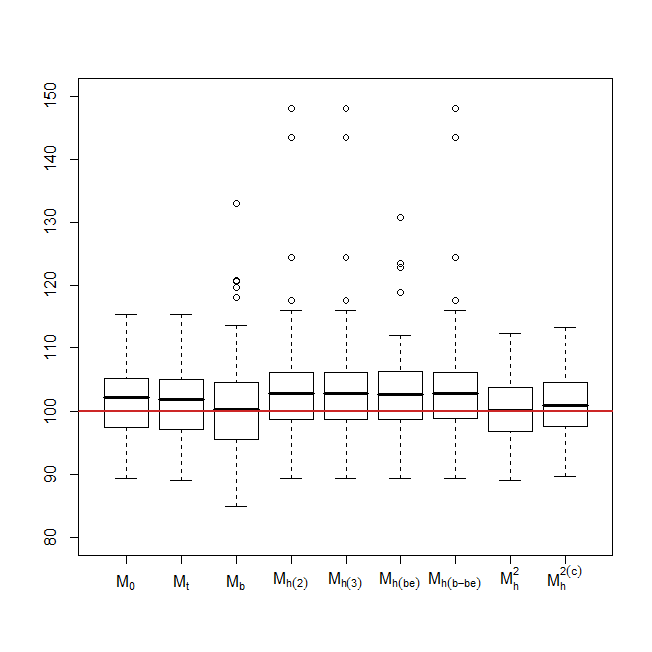}
\caption{Estimates of population size from nine models with two capture states for low mobility (left) and high mobility (right).  Parameter values used are given in the text.}
\label{fig:twostatescomp}
\end{center}
\end{figure}

\begin{figure}[!htp]
\begin{center}
\includegraphics[width=0.35\textwidth]{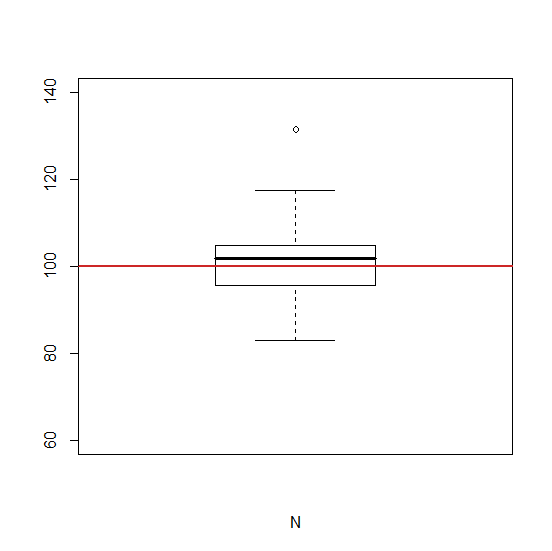}
\includegraphics[width=0.35\textwidth]{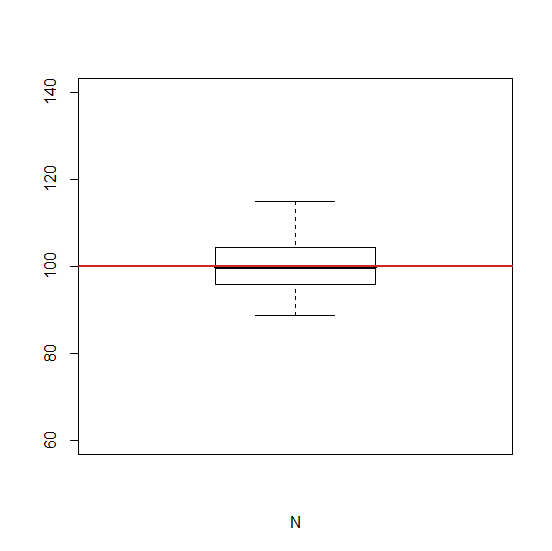}
\includegraphics[width=0.35\textwidth]{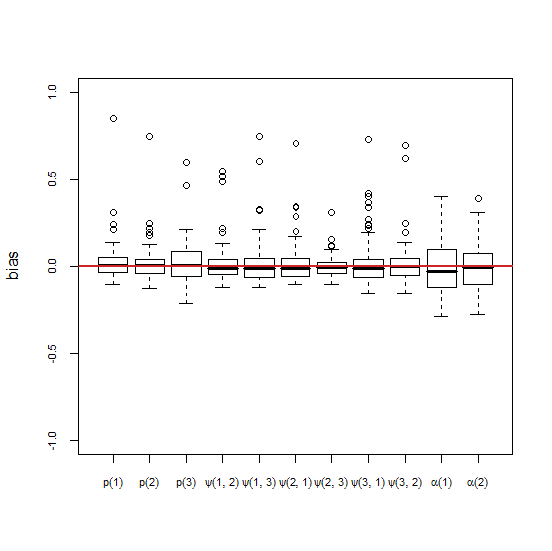}
\includegraphics[width=0.35\textwidth]{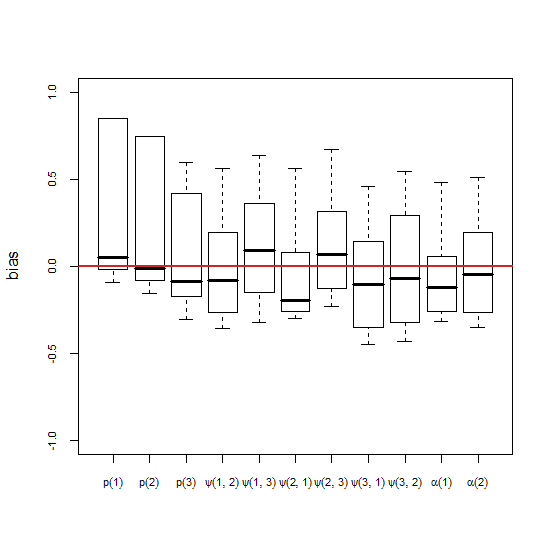}
\caption{Estimates of population size and bias of remaining model parameters for model $M_h^3$ for cases of low mobility (left) and high mobility (right).  Parameter values used are given in the text.}
\label{fig:threestates}
\end{center}
\end{figure}

\begin{figure}[!htp]
\begin{center}
\includegraphics[width=0.35\textwidth]{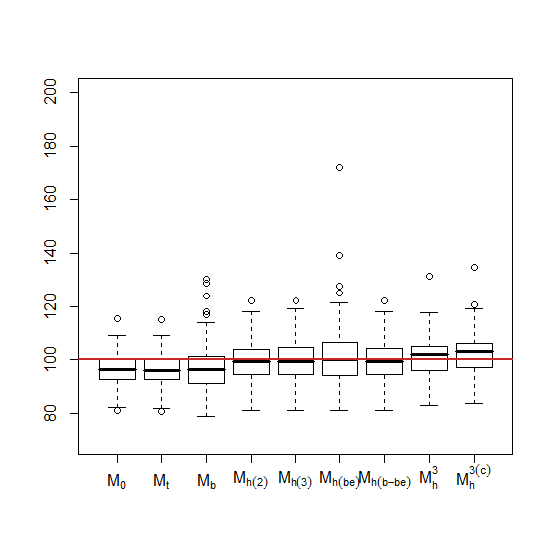}
\includegraphics[width=0.35\textwidth]{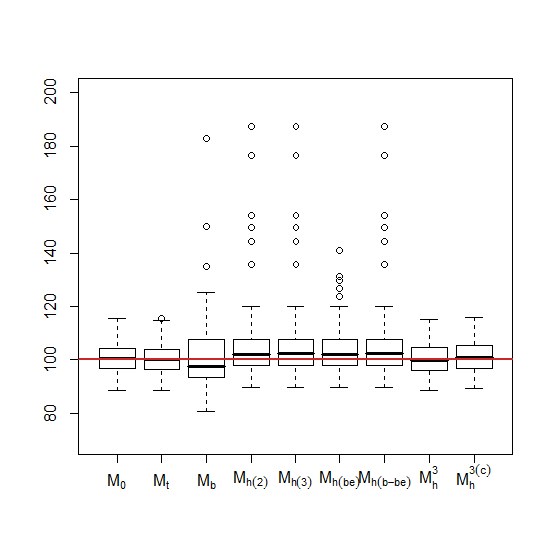}
\caption{Estimates of population size from eight models with three capture states for low mobility (left) and high mobility (right). Parameter values used are given in the text.}
\label{fig:threestatescomp}
\end{center}
\end{figure}

Figure~\ref{fig:twostates} shows boxplots of the estimates of $N$ and boxplots of the bias of the other model parameters for the true model $M_h^2$ ($R=2$).  Figure~\ref{fig:twostatescomp} displays boxplots of population size estimates from all the models when $R=2$. The corresponding plots for the three-state case are given in Figures~\ref{fig:threestates} and ~\ref{fig:threestatescomp}.

For the true model, in all cases considered, the estimates of $N$ do not show any bias.  In the two-state scenario the remaining model parameters are estimated well with no obvious differences in variation between low and high mobility.  In the three-state scenario the remaining model parameters are estimated accurately in the case of low mobility.  In the scenario of high mobility for the three-state case some of the remaining model parameter estimates appear to exhibit some bias and there is very large variation in all of the parameter estimates.  This appears to be due to an ``averaging'' or ``mixing'' effect across the states where there is greater uncertainty about the state of an individual when they are not captured leading to greater uncertainty in the parameter estimates.  The results from the traditional models without any individual heterogeneity indicate a strong negative bias for the $M_0$, $M_t$ and $M_b$ models in the case of low mobility for both the two- and three-state scenarios.  In the case of high mobility when there are two states the $M_b$ model shows no bias, whilst the $M_0$ and $M_t$ models show a positive bias.  When there are three states the $M_0$ and $M_t$ models show no bias whilst the $M_b$ model shows a slight negative bias.  For these three models the variability in the estimates of $N$ is similar for low and high mobility.  For the two-state low mobility scenario the heterogeneity models $M_{h(2)}$, $M_{h(3)}$, $M_{h(be)}$ and $M_{h(b-be)}$ all estimate $N$ with a slight negative bias suggesting that the mixtures are overestimating the different capture probabilities.  For the two-state high mobility scenario the heterogeneity models $M_{h(2)}$, $M_{h(3)}$, $M_{h(be)}$ and $M_{h(b-be)}$ appear to be positively biased to a similar extent. When there are two-states the heterogeneity models have similar precision for both low and high mobility.  In the three-state scenario for low mobility the heterogeneity models estimate $N$ reasonably well.  For high mobility they display a tendency to overestimate $N$ but only marginally.  When there are three-states the heterogeneity models have higher precision when there is high mobility.  In comparison to the existing heterogeneity models, the new multi-state model has the greatest precision of all the heterogeneity models considered for low and high mobility in both the two- and three-state cases.  The conditional model displays very similar results to the unconditional approach.

\section{Application - Great crested newts}\label{sec:newts}

These data are collected from a study site on the University of Kent campus.  Data have been collected on the population of newts breeding at the site since 2002.  Within this study capture occasions occur weekly throughout the breeding season, with individuals being identified through unique physical markings.  The additional state information corresponds to the pond in which the newts are captured.  Originally the site consisted of four ponds but was extended in 2009 to a total of eight ponds, four ``old'' ponds (state 1) and four ``new'' ponds (state 2), with the new ponds first being colonised in 2010.  Of specific interest is whether there were any differences between the old and new ponds in terms of capture and transition probabilities when they were first colonised and whether any differences have remained.  In order to assess this we compare results from the 2010 and 2013 data sets.  In order to make the assumption of closure reasonable we take a subset of six weeks ($T=6$) from the middle of each of the 2010 and 2013 breeding seasons, during which it can be assumed that all breeding newts have arrived at the breeding ponds and have not yet started to leave the area.

In the 2010 season a total of 33 unique individuals were encountered over the study period considered, whilst in 2013 there were 44 unique individuals encountered. We fit a range of models to the data, initially considering the new multi-state model with heterogeneity, model $M_h^2$.  Tables \ref{tab:Mhm22010} and \ref{tab:Mhm22013} provide the MLEs, standard errors (SEs) and 95\% confidence intervals (CIs) for the model parameters for  the 2010 and 2013 data. The  95\% CIs are calculated on a transformed scale for the parameters (logit or log scale) and back calculated to the raw scale to avoid intervals outside permissible boundary ranges.

\begin{table}
\begin{center}
\caption{\label{tab:Mhm22010} 2010 model $M_h^2$}
\begin{tabular}{|c|c|c|c|} \hline
Parameter   & MLE   & SE    & 95\% CI        \\ \hline
$N$         & 33.95 & 1.44  & (33.05, 51.23) \\
$p(1)$      & 0.82  & 0.083 & (0.60, 0.93)   \\
$p(2)$      & 0.33  & 0.045 & (0.25, 0.43)   \\
$\psi(1,2)$ & 0.31  & 0.076 & (0.18, 0.47)   \\
$\psi(2,1)$ & 0.03  & 0.019 & (0.01, 0.10)   \\
$\alpha(1)$ & 0.48  & 0.094 & (0.31, 0.66)   \\ \hline
\end{tabular}
\end{center}
\end{table}

\begin{table}
\begin{center}
\caption{\label{tab:Mhm22013} 2013 model $M_h^2$}
\begin{tabular}{|c|c|c|c|} \hline
Parameter   & MLE   & SE    & 95\% CI        \\ \hline
$N$         & 45.96 & 1.85  & (44.31, 56.50) \\
$p(1)$      & 0.36  & 0.061 & (0.25, 0.48)   \\
$p(2)$      & 0.41  & 0.048 & (0.32, 0.50)   \\
$\psi(1,2)$ & 0.05  & 0.035 & (0.01, 0.18)   \\
$\psi(2,1)$ & 0.08  & 0.033 & (0.03, 0.17)   \\
$\alpha(1)$ & 0.33  & 0.087 & (0.19, 0.52)   \\ \hline
\end{tabular}
\end{center}
\end{table}

\begin{table}
\begin{center}
\caption{\label{tab:Mt22013} 2013 model $M_t^2$}
\begin{tabular}{|c|c|c|c|} \hline
Parameter   & MLE   & SE    & 95\% CI        \\ \hline
$N$         & 45.63 & 1.69  & (44.21, 56.44) \\
$p_1$       & 0.39  & 0.074 & (0.26, 0.54)   \\
$p_2$       & 0.53  & 0.077 & (0.38, 0.67)   \\
$p_3$       & 0.35  & 0.072 & (0.23, 0.50)   \\
$p_4$       & 0.18  & 0.057 & (0.09, 0.31)   \\
$p_5$       & 0.46  & 0.076 & (0.32, 0.61)   \\
$p_6$       & 0.44  & 0.075 & (0.30, 0.59)   \\
$\psi(1,2)$ & 0.05  & 0.036 & (0.01, 0.19)   \\
$\psi(2,1)$ & 0.07  & 0.030 & (0.03, 0.16)   \\
$\alpha(1)$ & 0.32  & 0.082 & (0.18, 0.50)   \\ \hline
\end{tabular}
\end{center}
\end{table}

The MLEs of the capture probabilities indicate that in 2010 the old ponds had a higher capture probability than for the new ponds.  However, by 2013 the higher capture probability for the old ponds seems to have disappeared with similar capture probabilities for both old and new ponds (see below for discussion of model selection). The old ponds had more vegetation around the traps which meant that the newts had a greater chance of entering them than in the new ponds, where traps were more exposed.  In addition, for 2010 the transition probabilities indicate a general movement trend away from the old ponds to the new ponds, but once a newt reaches the new pond demonstrates a high fidelity to the new ponds. Previous analyses suggested that new recruits (first time breeders) used the new ponds more frequently than newts returning to the ponds \citep{lewis12}.  By 2013 the newts show high fidelity to both the old and new ponds. Finally we note that in 2010 the newts appear to be evenly distributed between the two ponds at the beginning of the study period, but by 2013, the proportion of newts increases in the new ponds by 2013 (though the confidence intervals are reasonably wide).

Interestingly, the results imply that only a single individual was missed during the study period in 2010 and two were missed during the 2013 study period.  Whilst the total population size is not the main focus of this study, these estimates are in keeping with the ecological understanding of the population.  It is believed that capture probability over the breeding season as a whole is very high. This was confirmed in 2005 when a drift fence was set up confirming that all individuals had been captured at least once.  A period of six weeks has been selected here in the central part of the breeding season, to accommodate the assumption of closure.  Outside of the selected six week period, in 2010, 7 individuals were seen before the selected period, but not during the study and one individual after but not during the study period.  No newts were captured both before and after.  Of those seen only before, all were seen quite early in the season, whilst the one individual seen after the study period is seen immediately after the 6 week period.  In 2013, 5 individuals were seen before, but not during, the closed period (all were seen very early in the season) and one individual is recaptured before and after the study period, but not during.

We now consider in further detail the issue of model selection. We fit the additional models $M_0^2$, $M_t^2$ and $M_{th}^2$ and compare them using the AIC statistic.  For model $M_{th}^2$ we specify the additional time dependence to be additive, so that $\logit \ p_t(2) = \logit \ p_t(1) + \eta$. Tables \ref{tab:AIC2010} and \ref{tab:AIC2013} provide the corresponding $\Delta$AIC values, estimates of $N$ and Pearson's chi-squared goodness-of-fit test for each model for both years of data. The model $M_h^2$ is deemed optimal for the 2010 data (only state dependent capture probabilities) and $M_t^2$ for the 2013 data (only time dependent capture probabilities). The corresponding parameter estimates for model $M_t^2$ for 2013 are provided in Table \ref{tab:Mt22013}. However, we note that in both cases the model $M_{th}^2$ has a $\Delta$AIC $<2$, indicating that there is little difference in support for the model deemed optimal and the model with both time and state dependent capture probabilities.

\begin{table}
\begin{center}
\caption{\label{tab:AIC2010} $\Delta$AIC values, MLEs and 95\% CIs for $N$ (denoted $\hat{N}$), Pearson $\chi^2$ statistics and corresponding $p$-values for the four multi-state models for the study on great crested newts for the 2010 data. A ``-'' denotes that the 95\% CI could not be calculated due to a boundary (or very near boundary) parameter estimate.}
\begin{tabular}{|c|c|cc|c|c|} \hline
Model      & $\Delta$AIC & $\hat{N}$ & (95\% CI) & $X^2$ & $p$-value \\ \hline
$M_0^2$    & 8.40 & 33.1 & - &  122.51 & $<$0.001 \\
$M_h^2$    & 0    & 34.0 & (33.1, 51.2) & 108.62 & $<$0.001\\
$M_t^2$    & 5.34 & 33.0 & - & 103.22 & $<$0.001\\
$M_{th}^2$ & 1.43 & 33.6 & (33.0, 68.0) & 91.62 & 0.005 \\ \hline
\end{tabular}
\end{center}
\end{table}

\begin{table}
\begin{center}
\caption{\label{tab:AIC2013} $\Delta$AIC values, MLEs and 95\% CIs for $N$ (denoted $\hat{N}$), Pearson $\chi^2$ statistics and corresponding $p$-values for the four multi-state models for the study on great crested newts for the 2013 data.  A ``-'' denotes that the 95\% CI could not be calculated due to a boundary (or very near boundary) parameter estimate.}
\begin{tabular}{|c|c|ccc|c|} \hline
Model      & $\Delta$AIC & $\hat{N}$ & (95\% CI) & $X^2$ & $p$-value \\ \hline
$M_0^2$    & 6.77 & 44.0 & - &  81.78 & 0.091 \\
$M_h^2$    & 6.67 & 46.0 & (44.3, 56.5) & 80.33 & 0.095 \\
$M_t^2$    & 0    & 45.6 & (44.2, 56.4) & 52.81 & 0.763 \\
$M_{th}^2$ & 1.68 & 45.7 & (44.2, 56.6) & 51.60 & 0.772 \\ \hline
\end{tabular}
\end{center}
\end{table}

All models fitted to the data suggest high fidelity to the old ponds in 2013 and the new ponds in both years with an increase in the proportion of individuals arriving at the new ponds in 2013 with similar estimates for the total population size,. The Pearson's chi-squared goodness-of-fit test does not indicate a lack of fit for the models fitted to the 2013 data. However, for the 2010 data, it is suggestive of a lack of fit. In conducting the goodness-of-fit test we do not pool small cells together so that the $p$-values will generally be an underestimate (i.e.~we are more likely to get a significant result than if an exact test is used). In addition, fewer individuals are observed in 2010 leading to an increase in the number of small cells observed, and hence an expectation of smaller $p$-values.

For comparative purposes, the estimates of population size resulting from alternative standard models are displayed in Tables \ref{tab:newtscomp2010} and \ref{tab:newtscomp2013} for data from 2010 and 2013 respectively. The models without any individual heterogeneity component ($M_0$, $M_t$ and $M_b$) all provide similar estimates for $N$. However, we note that for 2010, the estimate of $N$ is a boundary estimate for models $M_t$ and $M_b$. Further, for both 2010 and 2013 both the three group binomial mixture model, $M_{h(3)}$ and the binomial beta-binomial model $M_{h(b-be)}$ also lead to boundary estimates for model parameters, essentially reducing the model to the two binomial mixture model $M_{h(2)}$. For the 2013 data the estimates are generally similar to those obtained for the multi-state model. However, for the smaller dataset in 2010, the models without any individual heterogeneity appear to underestimate the population size and associated standard error, while the mixture models provide larger estimates (and very wide confidence intervals).

\begin{table}
\begin{center}
\caption{\label{tab:newtscomp2010} MLEs, standard errors (SE) and 95\% CIs for $N$ (denoted $\hat{N}$) for seven single-state models for the study on great crested newts for the 2010 data.  A ``-'' denotes that the 95\% CI could not be calculated due to a boundary (or very near boundary) parameter estimate.}
\begin{tabular}{|c|c|c|c|} \hline
Model         & $\widehat{N}$ & SE & 95\% CI \\ \hline
$M_0$         & 33.1 & 0.11 & (33.0, 33.6)  \\
$M_t$         & 33.0 & -   & -  \\
$M_b$         & 33.0 & - & -  \\
$M_{h(2)}$    & 41.5 & 17.39  & (33.2, 501.8) \\
$M_{h(3)}$    & 41.5 & -  & - \\
$M_{h(be)}$   & 34.7 & 2.45   & (33.1, 62.7)      \\
$M_{h(b-be)}$ & 41.5 & -   & -  \\ \hline
\end{tabular}
\end{center}
\end{table}

\begin{table}
\begin{center}
\caption{\label{tab:newtscomp2013} MLEs, standard errors (SE) and 95\% CIs for $N$ (denoted $\hat{N}$ for seven single-state models for the study on great crested newts for the 2013 data.  A ``-'' denotes that the 95\% CI could not be calculated due to a boundary (or very near boundary) parameter estimate.}
\begin{tabular}{|c|c|c|c|} \hline
Model         & $\widehat{N}$ & SE & 95\% CI \\ \hline
$M_0$         & 45.9 & 1.62 & (44.4, 52.1) \\
$M_t$         & 45.6 & 3.13 & (44.1, 62.6) \\
$M_b$         & 45.3 & 1.93 & (44.2, 54.8) \\
$M_{h(2)}$    & 46.5 & 2.45 & (44.3, 61.3) \\
$M_{h(3)}$    & 46.5 & -  & -              \\
$M_{h(be)}$   & 46.7 & 2.93 &   (44.3, 67.0)    \\
$M_{h(b-be)}$ & 46.5 & -  & -  \\ \hline
\end{tabular}
\end{center}
\end{table}

\section{Discussion}\label{sec:diss}

We have focussed on deriving multi-state closed capture-recapture models, where additional individual time-varying discrete covariates are observed. The models derived can be viewed as a closed population analogy to the AS model, assuming a first-order Markovian process for the transitions between states. The construction of an explicit closed-form (unconditional) likelihood via a set of sufficient statistics permits efficient evaluation of the likelihood and standard goodness-of-fit techniques, in the form of Pearson's chi-shared tests, to be applied. This can lead to generally small cell entries in the goodness-of-fit test, with different approaches for pooling cells and their interpretation a focus of current research.  Similarities of the closed multi-state model to the AS model also permit other extensions to be immediately applied. For example, in many cases, state may be only partially observed, including failure to observe a state when an individual is observed, or observing states with error \citep{king14}. Further, the modelling approach can be applied to the case of continuous individual time-varying covariates by considering an approximate (discretised) likelihood of multi-state form \citep{langrock13}.  The movement between the different states can also be generalised by removing the first-order Markov assumption, where the dwell-time distribution (the time spent in each state) is geometric, and instead imposing a more general dwell-time distribution, for example a shifted Poisson or negative-binomial distribution \citep{king16}. 

The proposed multi-state closed population model shows better accuracy and precision in estimating $N$ compared to competing models where the additional discrete state information is ignored. Further, additional insight can be obtained with regard to the states, which may themselves be of interest. Most notably, transition probabilities can be estimated (and hence the stable equilibrium distribution of the population over the states) and the relationship between state and capture probabilities evaluated. For the newt data analysis conducted, particular interest lay in the potential transition of newts from the old ponds to the new ponds installed in 2009 with a interest also in the total population size, not least with regard to the completeness of the data collection process and observing all individuals present. The analyses concluded that the data survey collection process appears to be close to a complete census of individuals present at the site which is unusual for capture-recapture studies. Further, that there was a general transition of newts from the old ponds to the new ponds between 2010 and 2013, but with little movement within the season. Finally, it appeared that there were initial differences between the capture probabilities in new and old ponds, in 2010, but once the new ponds had become established, the state-dependence was no longer significant by 2013.

In the presence of an underlying multi-state system process for closed populations, an unconditional likelihood can be derived and MLEs of the model parameters obtained, extending the previous conditional approaches. In the absence of the observed discrete covariate data, existing heterogeneity models appear to perform adequately, however, including the covariate information does improve the precision of the population estimate, as would be expected. The model developed here can be extended to the open population case, permitting both recruitment and departure from the study population, i.e.~to stopover models. This is the focus of current research.

\section*{Acknowledgements}

Worthington was funded by the Carnegie Trust for the Universities of Scotland and part funded by EPSRC/NERC grant EP/10009171/1 and McCrea by NERC fellowship grant NE/J018473/1. We would like to thank Amy Wright, Brett Lewis and the volunteer newt surveyors for collecting and collating the field data.  We would also like to thank Byron Morgan and Roland Langrock for helpful discussions regarding this work.

\end{document}